\documentclass[%
 reprint,
 amsmath,amssymb,
 aps,
]{revtex4-2}

\usepackage{graphicx}
\usepackage{dcolumn}
\usepackage{bm}
\usepackage{orcidlink} 

\begin{document}
\preprint{QgEUD}

\title{Quantum Generalized Equivalent Uniform Dose (QgEUD): A Simulation Method for Phase-Dependent Radiobiological Dose Effects}

\author{Yusuke Anetai\orcidlink{0000-0002-2284-7582}{$^{1, 2, *}$}  \thanks{E-mail: anetai.yus@kmu.ac.jp}}
 \affiliation{
 {$^{1}$Department of Radiology, Kansai Medical University, Osaka, Japan.}  \\
 {$^{2}$Division of Radiation Oncology, Kansai Medical University Hospital, Osaka, Japan.} 
 }

\begin{abstract}
The generalized equivalent uniform dose (gEUD) provides a biologically interpretable measure of heterogeneous dose distributions and is widely used in radiobiological modeling. However, because gEUD depends solely on dose magnitude, it does not explicitly account for collective cellular interactions or phase-dependent biological responses. Here, we propose a quantum generalized equivalent uniform dose (QgEUD), which extends the conventional gEUD kernel into the complex domain by introducing a phase variable while preserving the original dose-weighting formalism. This formulation yields a two-dimensional response surface that recovers conventional gEUD on the real axis and incorporates interaction-dependent radiobiological effects through phase modulation. The local response of the surface is characterized by a Kähler metric, providing an intrinsic measure of sensitivity to dose weighting and phase perturbations. To demonstrate the framework, local dose elements are modeled by an Ising Hamiltonian with dose- and phase-dependent interactions, and equilibrium response maps are obtained using Metropolis Monte Carlo simulations. Simulations in a virtual radiotherapy phantom preserve the overall dose distribution while producing spatially modulated biological-effect maps governed by collective interactions. The corresponding Kähler response identifies regions exhibiting enhanced sensitivity beyond dose magnitude alone, and parameter sensitivity analysis confirms stable convergence under practical simulation conditions. These results establish QgEUD as a quantum-inspired extension of gEUD that integrates heterogeneous dose aggregation, phase-dependent interactions, and geometric response within a unified mathematical framework, providing a basis for interaction-aware radiobiological modeling and future quantum-compatible optimization.
\end{abstract}


\maketitle

\section{\label{sec:level1} Introduction \protect }

Radiotherapy aims to deliver a sufficient radiation dose to the tumor while minimizing exposure to surrounding normal tissues. Treatment planning is conventionally based on voxel-wise dose calculations and evaluated using dosimetric indices such as dose–volume histograms (DVHs), mean dose, maximum dose, and related metrics \cite{refs01,refs02,refs03, refs03b, refs03c}. However, identical physical doses may produce substantially different biological responses owing to variations in cell type, oxygenation status, cell-cycle phase, DNA repair capacity, and intrinsic radiosensitivity \cite{refs04,refs05,refs06,refs07}. Furthermore, when dose heterogeneity and intercellular interactions are considered, physical dose alone may not adequately characterize the biological response of irradiated tissues.

The generalized equivalent uniform dose (gEUD) was introduced to summarize heterogeneous dose distributions into a biologically meaningful scalar quantity using a generalized power mean \cite{refs08}. By adjusting the tissue-specific parameter $a$, gEUD preferentially emphasizes low-, intermediate-, or high-dose regions and has become a widely used tool for radiobiological dose evaluation and treatment-plan optimization. Despite its practical utility, however, gEUD remains fundamentally dose-based and does not explicitly represent spatial correlations, collective cellular interactions, or phase-dependent biological responses.

A more comprehensive description of radiobiological effects requires simultaneous consideration of heterogeneous dose distributions, stochastic energy deposition, DNA damage and repair, cellular state transitions, and intercellular interactions \cite{refs06, refs07,refs09, refs10, refs11}. Such problems naturally lead to many-body and combinatorial optimization formulations, for which exhaustive classical computation rapidly becomes intractable as the number of cellular states increases \cite{refs12, refs13}. Because these formulations can be represented by Ising Hamiltonians or quadratic unconstrained binary optimization (QUBO) models, they are naturally compatible with quantum-inspired optimization algorithms and future quantum-computing architectures \cite{refs13, refs15, refs16}.

Motivated by these considerations, we propose a quantum generalized equivalent uniform dose (QgEUD) as a phase-extended formulation of conventional gEUD. In the framework, dose-derived information is treated as an amplitude component, whereas an additional phase variable encodes relative cellular interactions and radiation-response relationships. This enables QgEUD to preserve the dose-aggregation capability of gEUD while representing interaction-dependent enhancement, suppression, and interference-like effects in a unified mathematical framework. In this work, we formulate the QgEUD framework, introduce a Kähler-geometric interpretation of its response manifold, and demonstrate its feasibility using a Metropolis-based Ising simulation in a virtual radiotherapy phantom. The proposed approach provides a proof of concept for phase-sensitive, interaction-aware radiobiological dose evaluation compatible with future quantum-inspired and quantum-computing methodologies.


\section{\label{sec:level1} Materials and Methods \protect}
\subsection{\label{sec:level2} Equivalent uniform dose metrics for radiobiological effects}
To account for the biological effects of heterogeneous dose distributions, the gEUD was employed in this study. The gEUD represents a non-uniform dose distribution by the uniform dose that would produce an equivalent biological response. Owing to this property, gEUD has been widely used as an intermediate dosimetric quantity for radiobiological modeling, including tumor control probability (TCP)  and normal tissue complication probability (NTCP)  estimation \cite{refs17, refs18}.

The gEUD is defined as:
\begin{equation}
\mathrm{gEUD} = \left( \sum_{j} v_j {D_j}^a \right)^{1/a}
\label{eq:1},
\end{equation}
where $D_{j}$ denotes the dose delivered to voxel $j$, $v_j$ is the corresponding fractional volume, and $a$ is a tissue-specific volume-effect parameter. Positive values of $a$ emphasize high-dose regions and are typically used for normal-tissue evaluation, whereas negative values of $a$ increase the influence of low-dose regions and are often applied to target-volume assessment and TCP-related analyses.
The gEUD concept was originally developed from the equivalent uniform dose (EUD) framework proposed for radiobiological response modeling \cite{refs08}. In this framework, a heterogeneous dose distribution is converted into a biologically equivalent uniform dose, which can subsequently be incorporated into tumor control probability (TCP) and normal tissue complication probability (NTCP) models. For TCP modeling, a logistic formulation is commonly adopted \cite{refs17b, refs17c}:

\begin{equation}
\mathrm{TCP} = \frac{1}{1+ \left(  \frac{\mathrm{TCD}_{50}}{\mathrm{gEUD}}  \right)^{4\gamma_{50}} }
\label{eq:2},
\end{equation}
where $\mathrm{TCD}_{50}$ denotes the dose yielding 50\% tumor control and $\gamma_{50}$ represents the normalized dose–response gradient at $\mathrm{TCD}_{50}$.

Similarly, NTCP can be expressed using the Lyman–Kutcher–Burman (LKB) formalism \cite{refs18, refs18d}:
\begin{equation}
\mathrm{NTCP} = \mathrm{erf} \left( \frac{\mathrm{gEUD}-\mathrm{TD}_{50}}{m \cdot \mathrm{TD}_{50}} \right)
\label{eq:3},
\end{equation}
where $\mathrm{erf}(\cdot)$ is the error function, $\mathrm{TD}_{50}$ is the tolerance dose associated with a 50\% complication probability, and $m$ characterizes the steepness of the dose–response relationship. The volume-effect parameter $a$ in the gEUD formalism is related to the volume parameter $n$ in the LKB model through $a=1/n$ \cite{refs18, refs18b, refs18c}. 

Consequently, tissues exhibiting serial-organ behavior correspond to small values of $n$ and therefore large positive values of $a$, causing gEUD to be dominated by high-dose regions. In contrast, parallel organs exhibit larger values of $n$ and smaller values of $a$, resulting in a stronger volume-averaging effect. This reciprocal relationship allows gEUD to provide a biologically interpretable bridge between heterogeneous dose distributions and TCP/NTCP-based radiobiological outcome modeling.

\subsection{\label{sec:level2} Quantum generalized equivalent uniform dose}
The gEUD summarizes a heterogeneous dose distribution as a single biologically equivalent uniform dose and has been widely incorporated into TCP and NTCP models. However, because the gEUD is derived solely from voxel-wise dose magnitudes, it does not explicitly account for spatial correlations or collective interactions among irradiated cells. To incorporate these features, we introduce a quantum-inspired extension of gEUD, termed the quantum generalized equivalent uniform dose (QgEUD).

For a dose distribution consisting of dose bins $D_j$ with fractional volumes $v_j$, a normalized wave function was defined as:
\begin{equation}
| \psi_j \rangle := \sqrt{\frac{v_j}{\sum_{k}v_k}} | j \rangle \,,  \sum_{j} \left |\psi_{j} \right|^2 = \langle \psi | \psi  \rangle = 1
\label{eq:4},
\end{equation}
where $| j \rangle$ denotes the basis state associated with dose element $j$. This formulation allows the differential dose–volume histogram $\mathrm{dDVH} := v$ to be interpreted as a probability amplitude distribution, where $\left| \psi_j \right|^2$ represents the probability of observing dose level $D_j$. In this framework, the gEUD is represented as a quantum-like state,
\begin{equation}
\mathrm{gEUD}(a) = { \langle \psi | \hat{D}^a_{j} | \psi \rangle }^{1/a}
\label{eq:5},
\end{equation}
where $\hat{D}$ represents the diagonal dose operator with eigenvalues $D_j$.
To incorporate biological phase information $\theta$, a complex phase factor $e^{i\theta}$ can be assigned to each dose state. In this study, QgEUD was defined by extending the dose kernel of the conventional gEUD into the complex domain. 
\begin{equation}
\mathrm{gEUD}(z) = { \langle \psi | \hat{D}^z_{j} | \psi \rangle }^{1/z}
\label{eq:6}.
\end{equation}
Here ${z = a+i\theta}$ is a complex-valued parameter extending the conventional gEUD formalism from $\mathbb{R}$ to $\mathbb{C}$, where $a$ denotes the conventional gEUD $a$-factor and ${\theta}$ denotes the phase parameter. 

However, a geometric analysis presented later reveals that the limit at $z=0$ is not well-defined when the exponent is also extended to the complex domain. To preserve continuity and maintain a unique geometric interpretation around the singular point, only the dose kernel was extended into the complex domain, while the outer exponent was retained as the real-valued parameter $a$. Furthermore, to ensure dimensional consistency of the complex logarithmic phase term, an arbitrary reference dose $D_{\mathrm{ref}}$ was introduced so that the quantity $(D/D_{\mathrm{ref}})$ became dimensionless. In this study, $D_{\mathrm{ref}}$ was defined as the maximum dose value in the two-dimensional dose distribution $D_{jk}$, i.e., $D_{\mathrm{ref}}:= \mathrm{max}(D_{jk})$. Accordingly, the QgEUD was finally defined as:
\begin{equation}
\mathrm{QgEUD}(a, \theta) = D_{\mathrm{ref}}\left( { \left\langle \psi \right| \left( \frac{\hat{D}}{D_\mathrm{ref}} \right)^{a+i \theta}  \left| \psi \right\rangle } \right) ^{1/a}
\label{eq:7}.
\end{equation}

This interpretation further suggests that the point $\theta = 0$ and $a \rightarrow 0$ acts as a geometric singularity connecting the positive- and negative-$a$ branches of the QgEUD manifold.

\subsection{\label{sec:level2} Geometrical characteristic of QgEUD}
The real-valued QgEUD surface is defined by:
\begin{equation}
\Omega(a,\theta) := \left| \mathrm{QgEUD}(a, \theta) \right|
\label{eq:8}.
\end{equation}
The conventional gEUD curve is recovered on the real section ${\theta=0}$, namely,
\begin{equation}
\Omega(a,0) :=  \mathrm{gEUD}(a)
\label{eq:9}.
\end{equation}
Therefore, QgEUD can be regarded as a two-dimensional phase-extended manifold whose real-axis section corresponds to the classical gEUD curve \cite{refs19a, refs19b, refs19c}. The gEUD curve $\Omega(a,0)$ is a monotonically increasing sigmoid-shaped function with two vertices and a single inflection point, and is bounded by $0 \leq D_\mathrm{min} \leq \Omega(a,0) \leq D_\mathrm{max}$ \cite{refs19b, refs19c}. Geometrically, the QgEUD surface can be interpreted as a two-sheeted funnel-like structure connected at the singular point ${a=0,\theta=0}$, and  bounded above by the $D_\mathrm{max}$-plane and below by the $D_\mathrm{min}$-plane. The positive-$a$ sheet corresponds to high-dose-weighted behavior, whereas the negative-$a$ sheet corresponds to low-dose-weighted behavior. The phase parameter $\theta$ induces oscillatory deformation around these sheets, producing a wavy surface structure. 

This phase-extended geometry provides a mathematical framework for incorporating spatial or biological collective effects into gEUD-based dose evaluation. Whereas conventional gEUD represents dose heterogeneity only through voxel-wise dose magnitudes \cite{refs20a, refs20b}, QgEUD additionally allows interference-like or coherence-like effects to be encoded through the phase parameter $\theta$. Thus, the QgEUD manifold generalizes the gEUD curve from a real dose-response curve to a complex, phase-sensitive dose-response surface (Fig. 1).

\begin{figure}[b]

\includegraphics[width=0.95\columnwidth]{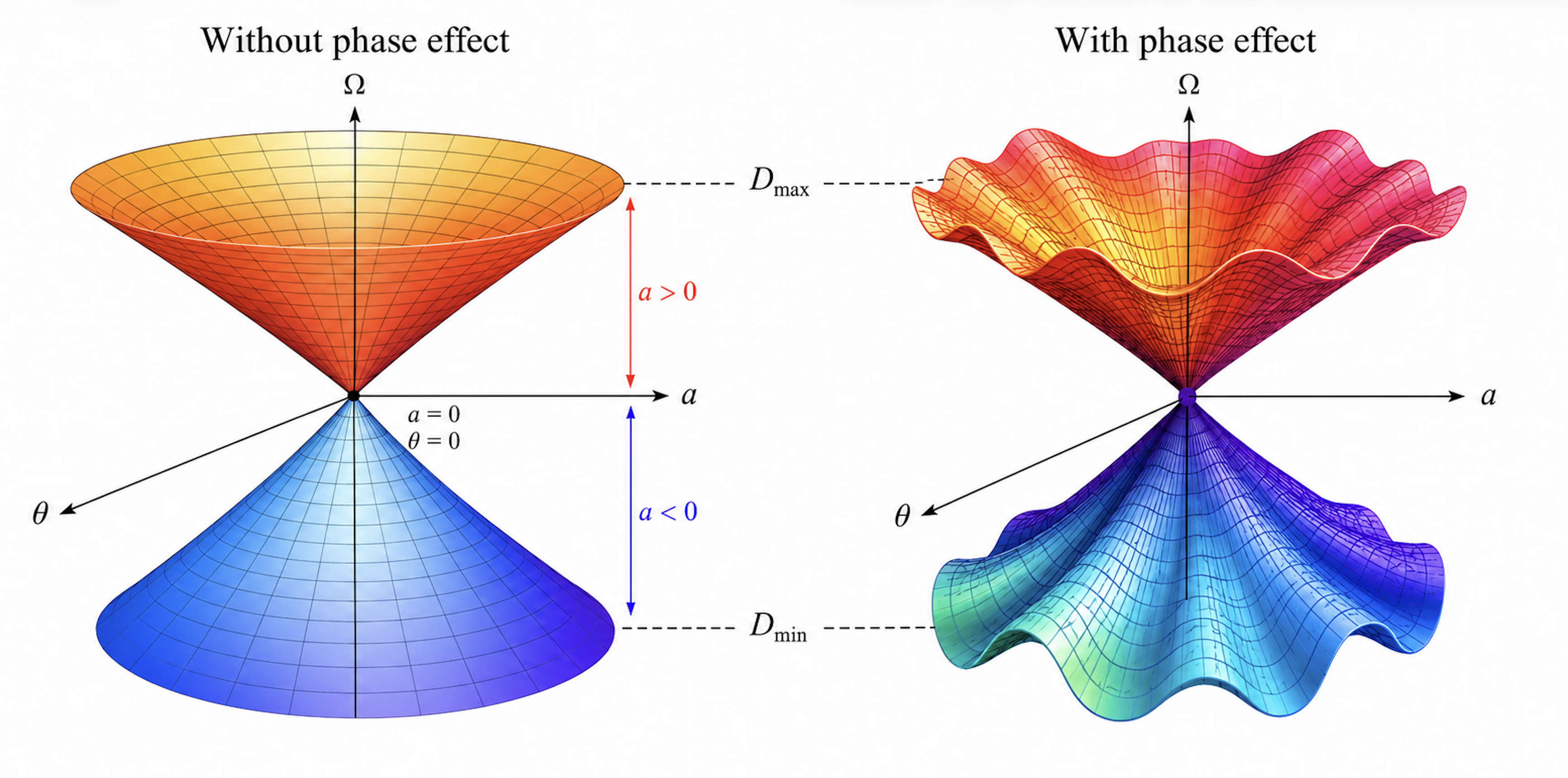}

\caption{\label{fig:epsart} 
Geometric representation of the QgEUD manifold with and without phase modulation. The QgEUD surface is illustrated as a two-sheeted funnel-like manifold defined over the real gEUD exponent $a$ and the phase parameter $ {\theta} $, with the vertical axis representing $\Omega = QgEUD$. (a) In the absence of a phase effect, the manifold forms smooth upper and lower sheets connected at the singular point $a=0, \theta=0$, where the conventional geometric-mean limit is approached. The upper sheet corresponds to the positive-$a$ branch, which emphasizes high-dose contributions and approaches $D_\mathrm{max}$, whereas the lower sheet corresponds to the negative-$a$ branch, which emphasizes low-dose contributions and approaches $D_\mathrm{min}$. (b) When the phase effect is introduced, oscillatory modulation along the $\theta$ direction deforms both sheets into wavy funnel-like surfaces. This deformation represents phase-dependent interference or coherence effects in the QgEUD formulation, while preserving the connection of the positive- and negative-$a$ branches at the singular point.
}
\end{figure}

Now, we introduce a kernel of QgEUD:
\begin{equation}
	\begin{split}
		K(z) &:= \sum_{j} v_{j} \left( \frac{D_{j}}{D_\mathrm{ref}} \right)^{z} \\
			&= \sum_{j} v_{j} \left( \frac{D_{j}}{D_\mathrm{ref}} \right)^{a} e^{i\theta \mathrm{log} (D_j / D_{\mathrm{ref}})}.
	\end{split}
\label{eq:10}
\end{equation}
Therefore,
\begin{equation}
	\Omega(z) := D_{\mathrm{ref}}|K(z)|^{1/a}
\label{eq:11}.
\end{equation}
In this formulation, the holomorphic gEUD kernel $K$ serves as a complex generating function and is analogous to a partition function. Introducing the parameter $x_j = \mathrm{log} (D_j/D_{\mathrm{ref}})$, the first derivative of $\mathrm{log}K$ with respect to the relevant generating parameter $z$ yields the expectation value of $x_j$:
\begin{equation}
	\begin{split}
		\partial_{z} \mathrm{log} K &= \sum_{j} p_j(z)x_j = \langle x \rangle, \\
		p_j(z) &= \frac{v_j e^{zx_j}}{\sum_{k}v_k e^{z x_k}} = \frac{v_j e^{zx_j}}{K} ,
	\end{split}
\label{eq:12}
\end{equation}
where $p_j$ denotes the relative contribution of the $j$-th logarithmic dose bin. Moreover, the second derivative of $\mathrm{log}K$ with respect to the generating parameter $z$ yields the variance of the logarithmic dose.
\begin{equation}
	\partial_{z}^2 \mathrm{log} K =\langle x^2 \rangle - \langle x \rangle^2 = \mathrm{Var}(x)
\label{eq:13},
\end{equation}
where $\mathrm{Var}(\cdot)$ is variance.
Thus, we can introduce a Kähler metric as an intrinsic measure derived from a potential function on the complex parameter space, thereby integrating the dose-weighting and phase-modulation directions into a single local geometric quantity. The Kähler potential ${\Phi}$ is defined as:

\begin{equation}
\Phi(a,\theta) :=  \mathrm{log} \left| \frac{\Omega}{D_{\mathrm{ref}}} \right|^2 = \frac{1}{a} \mathrm{log} \left| K \right|^2
\label{eq:14}.
\end{equation}
Then, the Kähler metric $g_{z\bar{z}}^{(\Omega)}$ is given as:
\begin{equation}
	\begin{split}
		g_{z\bar{z}}^{(\Omega)} &= \partial_{z} \partial_{\bar{z}}\Phi = \frac{1}{4} \left( {\partial}^2_{a} + {\partial}^2_{\theta} \right) \Phi \\
				   &= \frac{1}{a^2} \left( \frac{1}{a}\mathrm{log}|K| - {\partial}_a \mathrm{log}|K| \right).
	\end{split}
\label{eq:15}
\end{equation}
By expanding Equation (15), we obtain:
\begin{equation}
g_{z\bar{z}}^{(\Omega)} = -\frac{\beta}{a\Omega}.
\label{eq:16}
\end{equation}
Here $\beta:= {\partial}_a \Omega$ denotes the local gradient of the QgEUD surface with respect to $a$. Since $\Omega \geq D_{\mathrm{min}} \geq 0$ and $\beta \geq 0$ for the monotonically increasing gEUD curve, the metric $g_{z\bar{z}}^{(\Omega)}$ is not necessarily positive definite over the entire QgEUD surface. Therefore, in this study, $g_{z\bar{z}}^{(\Omega)}$ is interpreted as a pseudo-Kähler metric.

The metric $g_{z\bar{z}}^{(\Omega)}$ characterizes the local geometric response of the QgEUD surface. In terms of surface deformation, the QgEUD surface becomes geometrically more rigid as $|a|$ increases, whereas regions with large $\beta$ or small $|a|$ exhibit greater sensitivity to the external force that bends or deforms the surface. At the same time, the dose leverage response \cite{refs19c} increases with $|a|$. The sign of the metric reflects the reversal of the surface response between the positive- and negative-$a$ funnel-like sheets, indicating that the two branches respond oppositely to perturbations in the dose-weighting direction.
The pseudo-Kähler metric also evaluates the second-order local response of the QgEUD potential in both the real dose and phase directions. Thus, $g_{z\bar{z}}^{(\Omega)}$ acts as a unified local sensitivity measure of the QgEUD surface. Now, the line element ${ds}$, which encodes both the magnitude and orientation of the local response on the QgEUD surface, is given by
\begin{equation}
ds^2 = 2g_{z\bar{z}}^{(\Omega)} dzd\bar{z} = 2g_{z\bar{z}}^{(\Omega)} (da^2+d{\theta}^2)
\label{eq:17}.
\end{equation}
Given an infinitesimal displacement $\delta a$ and $\delta \theta$, the corresponding metric energy $ \delta E$ is expressed as
\begin{equation}
\delta E = \frac{1}{2} \mathfrak{H}_{K} \left( \delta {a}^2 + \delta {\theta}^2 \right)
\label{eq:18},
\end{equation}
where ${\mathfrak{H}_K}$ denotes the geometric response (or sensitivity) required to induce an infinitesimal displacement along the ${a}$- or ${\theta}$-direction on the QgEUD surface. From Equation (17), it is naturally defined as:
\begin{equation}
\mathfrak{H}_{K} = 2\left| g_{z\bar{z}}^{(\Omega)} \right|
\label{eq:19}.
\end{equation}
A high geometric response indicates strong sensitivity to changes in the $a$-value, phase modulation, and intercellular interactions, suggesting that the resulting biological effect is less directly determined by the dose magnitude alone. In contrast, a low geometric response indicates that the expected effect is more closely governed by the local dose value.

\subsection{\label{sec:level2} Biological phase aggregation}
The phase parameter $\theta$ in QgEUD was interpreted as an effective collective coordinate representing multiple unresolved biological factors, including oxygenation, cell-cycle status, DNA repair capacity, chromatin state, and intercellular signaling. Here, a normalized biological state is introduced as:
\begin{equation}
| \Psi (\bm{\lambda}) \rangle = \frac{\sum_{j} \sqrt{v_j} \mathrm{exp}\left[ \frac{1}{2} {\bar{\eta}}_j (\bm{\lambda}) + \frac{i}{2} {\bar{\xi}}_j (\bm{\lambda}) \right]}{\sqrt{\sum_{j} v_j e^{{\bar{\eta}}_j (\bm{\lambda})}}} | j \rangle
\label{eq:20},
\end{equation}
where the parameter vector $\bm{\lambda}$ collectively specifies the coordinates of the state $| \Psi (\bm{\lambda}) \rangle$, comprising the dose-response parameters $a_r$ and biological phase variables ${\theta}_r$, as defined below.
\begin{equation}
\bm{\lambda} = (a_1, \cdots, a_m, {\theta}_1, \cdots, {\theta}_{m})^{\top}
\label{eq:21}.
\end{equation}
Here, $\bar{\eta}_j(\bm{\lambda} )$ and $\bar{\xi}_j(\bm{\lambda})$ denote the aggregated biological response associated with $a$-dependent amplitude and phase-dependent contributions, respectively, for the $j$-th cellular element.
\begin{equation}
	\begin{split}
		\bar{\eta}_j &= \sum_{r}  a_r \eta_{jr} \\
		\bar{\xi}_j &= \sum_{r}  \theta_r \xi_{jr} , 
	\end{split}
\label{eq:22}
\end{equation}
where $\eta_{jr}$ and $\xi_{jr}$ represent the factor-specific biological response functions associated with $r$-th biological component. The normalized probability assigned to the $j$-th cellular element is then defined as:
\begin{equation}
p_j (\bm{\lambda}) = \frac{v_j \mathrm{exp}({\bar{\eta}_j(\bm{\lambda})})}{\sum_{j} v_j \mathrm{exp}({\bar{\eta}_j(\bm{\lambda})})}
\label{eq:23}.
\end{equation}
Throughout this study, the state $| \Psi (\bm{\lambda}) \rangle$ is assumed to satisfy $ \langle \Psi (\bm{\lambda}) |\Psi (\bm{\lambda}) \rangle = 1$.

Focusing on a cellular dose element $j$, the effective phase can be defined as:
\begin{equation}
\bar{\theta}_{j} \sim {\sum}^{m}_{r} f({\theta}_{jr}){\theta}_{jr}
\label{eq:24},
\end{equation}
where $f({\theta}_{jr})$ denotes the relative contribution of each biological factor ${\theta}_{jr}$. Thus, $\bar{\theta}_j$ represents a coarse-grained aggregation of multiple biological effects rather than their mechanistic separation. 
Because the individual biological phases were not directly measurable, their combined variation was represented statistically as
\begin{equation}
\bar{\theta}_{j} \sim p_{\mathrm{eff}} (\theta)
\label{eq:25}.
\end{equation}
The effective distribution $p_{\mathrm{eff}}$ corresponds to the circular convolution of the individual phase distributions with angular scaling and summarizes the combined uncertainty arising from multiple unobserved biological factors. Thus, random phase assignment provides a statistically consistent ensemble representation of biological heterogeneity when factor-specific measurements are unavailable. This aggregation is appropriate for the present objective because the calculation does not seek to identify the independent contribution of each biological mechanism. Rather, it estimates the collective response field produced by their combined phase-dependent interactions. 

\subsection{\label{sec:level2} Quantum-geometric equilibrium representation}
The local geometry of the parameterized response states can be evaluated using the quantum geometric tensor,
\begin{equation}
Q_{\mu \nu} = \langle \partial_{\mu} \Psi | \left(1- | \Psi \rangle \langle \Psi |\right) |\partial_{\nu} \Psi \rangle 
\label{eq:26},
\end{equation}
where $\mu, \nu$ are tensor indices, and $\partial_{\mu} = \partial/\partial {\lambda}^{\mu}$. Its real and imaginary parts are given by:
\begin{equation}
g_{\mu \nu} = \mathrm{Re}\left( Q_{\mu \nu} \right), \, F_{\mu \nu} = -2\mathrm{Im} \left(Q_{\mu \nu} \right)
\label{eq:27}.
\end{equation}
The quantum metric $g_{\mu\nu}$ provides a quantitative measure of the local distinguishability and sensitivity of the response state under variations in $a$, $\theta$, and other biological parameters. The Berry curvature $F_{\mu \nu}$ quantifies the geometric phase accumulated when these parameters are varied along a closed path. The diagonal metric components represent sensitivity to individual parameters, whereas the off-diagonal components describe coupled or cross-parameter sensitivity. 

Now, the corresponding macroscopic equilibrium state $\Psi_{eq} (\bm{\lambda})$ can be represented using real, non-negative square-root amplitudes as
\begin{equation}
|\Psi_{\mathrm{eq}} (\bm{\lambda}) \rangle = \sum_{s} \sqrt{P_{\mathrm{eq}}(s | \bm{\lambda})} |s \rangle
\label{eq:28},
\end{equation}
where $P_{\mathrm{eq}}(s | \bm{\lambda})$ denotes the equilibrium probability distribution when a spin configuration $s \in \{-1, +1\}$ is specified by the parameter vector $\bm{\lambda}$:

\begin{equation}
	\begin{split}
		P_{\mathrm{eq}}(s | \bm{\lambda}) &= \frac{1}{Z(\bm{\lambda})} \mathrm{exp} \left(-\frac{H(s | \bm{\lambda})}{T} \right), \\ 
		Z(\bm{\lambda}) &:= \sum_{s} \mathrm{exp} \left(-\frac{H(s | \bm{\lambda})}{T} \right) \, ,
	\end{split}
\label{eq:29}
\end{equation}
where $T$ denotes temperature, $H$ is the Hamiltonian, and $Z$ is the corresponding partition function. Accordingly, $P_{\mathrm{eq}}(s | \bm{\lambda})$ follows the Boltzmann distribution.
Here, the Berry connection trivially vanishes in the chosen equilibrium gauge.
\begin{equation}
A_{\mu} = i \langle \Psi_{\mathrm{eq}} | \partial_{\mu}  \Psi_{\mathrm{eq}} \rangle = \frac{i}{2} \partial_{\mu} \sum_{s}  P_{\mathrm{eq}} (s | \bm{\lambda}) = 0
\label{eq:30}.
\end{equation}
Therefore, Berry curvature of the equilibrium state is given by
\begin{equation}
F_{\mu \nu}^{\mathrm{eq}} = \partial_{\mu} A_{\nu} - \partial_{\nu} A_{\mu} = 0
\label{eq:31}.
\end{equation}
This Berry-flat condition indicates that the equilibrated macroscopic response is free from any local path-dependent geometric phase and can therefore be treated as a state function of $\bm{\lambda}$. Accordingly, the quantum geometric tensor is determined by the quantum metric:
\begin{equation}
	\begin{split}
		g_{\mu \nu}^{\mathrm{eq}} &=  \mathrm{Re} (\langle \partial_{\mu} \Psi_{\mathrm{eq}} | \partial_{\nu} \Psi_{\mathrm{eq}}\ \rangle) \\
			&= \sum_{s} P_{\mathrm{eq}} (\partial_{\mu} \mathrm{log} P_{\mathrm{eq}}) ( \partial_{\nu} \mathrm{log} P_{\mathrm{eq}}) \\
			&= \frac{1}{4T^2}\mathrm{Cov}(\partial_{\mu} H, \partial_{\nu}H),
	\end{split}
\label{eq:32}
\end{equation}
where $\mathrm{Cov}(\cdot, \cdot)$ is covariance, and the metric is directly related to the Fisher information metric. 

Analogously, the quantum metric associated with QgEUD can be derived from the normalized probability distribution defined in Eq. (12):
\begin{equation}
g_{z\bar{z}}^{(Q)} = \frac{1}{4}{\partial_{a}}^2 \mathrm{log}K = \frac{1}{4} \mathrm{Var}(x) > 0
\label{eq:33}.
\end{equation}
It therefore establishes a natural metric connection between the phase-extended microscopic metric $(g_{z\bar{z}}^{(Q)})$ and the macroscopic biological-effect metric $(g_{z\bar{z}}^{(\Omega)})$ within the QgEUD framework.
\begin{equation}
g_{z\bar{z}}^{(Q)} = -\frac{1}{4a} \frac{\partial}{\partial a} \left( a^3 g_{z\bar{z}}^{(\Omega)} \right)
\label{eq:34}.
\end{equation}

\subsection{\label{sec:level2} Connection between Microscopic Biological Interactions and the Macroscopic QgEUD Field}
The microscopic holomorphic gEUD kernel $K(\bm{\lambda})$ defines the normalized state $\Psi(\bm{\lambda})$ through the probability distribution $p_j(\bm{\lambda})$.
The resulting microscopic state is subsequently incorporated into the interaction-dependent Hamiltonian, which determines the equilibrium probability distribution $P_{\mathrm{eq}} (s | \bm{\lambda})$.
Finally, this distribution defines the macroscopic equilibrium state $\Psi_{\mathrm{eq}}(\bm{\lambda})$, whose vanishing Berry curvature implies the absence of local path-dependent geometric-phase effects.

\subsection{\label{sec:level2} QgEUD-based biological effect simulation using a metropolis-based Ising model}
To demonstrate QgEUD-based estimation of biological effects incorporating phase-dependent interactions, we applied an Ising-type model to the calculation \cite{refs12, refs13, refs21}. In this study, each 1-pixel $(2 \, \mathrm{mm} \times 2 \, \mathrm{mm})$ region was regarded as a cellular dose element, to which a phase and an $a$-value were assigned. 

Each cellular element was assigned a spin state, $s_{jk} \in \{ -1,+1\}$, where the spin state represents the local phase state or interaction polarity of the corresponding dose component. The system energy was defined using the following Ising Hamiltonian:
\begin{equation}
H = -\sum_{\langle jk,lm \rangle}  J_{jk,lm} s_{jk} s_{lm} - \sum_{\langle j,k \rangle} h_{jk}s_{jk}
\label{eq:35}.
\end{equation}
Here, $J_{jk,lm}$ denotes the coupling coefficient between four-neighboring cellular elements $(j,k)$ and $(l,m)$, and is defined as:
\begin{equation}
J_{jk,lm} = w_{J} J_{0} \mathrm{exp} \left( -\frac{\left| D^{a_{jk}}_{jk} - D^{a_{lm}}_{lm} \right| } {\sigma_{D}} \right) \mathrm{cos} \left( \theta_{jk} - \theta_{lm} \right)
\label{eq:36},
\end{equation}
where $D_{jk}$ and $D_{lm}$ are the local doses assigned to the corresponding cellular elements, $\theta_{jk}$ and $\theta_{lm}$ are their phase parameters, $a_{jk}$ and $a_{lm}$ are the gEUD $a$-value factors assigned at each cell, and $\sigma_D$ controls the sensitivity of the coupling strength to dose differences, which is defined as standard deviation of dose distribution $D^{a_{jk}}_{jk}$. $J_0$ was introduced as a reference coupling constant that defines the intrinsic interaction scale of the Ising system before dose- and phase-dependent modulation. The weighting factor $w_J$ is a hyperparameter that determines the relative contribution of the interaction term to the system Hamiltonian in the QgEUD-based simulation. This formulation assigns stronger coupling to neighboring elements with similar dose values, while the cosine term modulates the interaction according to their relative phase difference.

The local external field $h_{jk}$, which incorporates local dose information into the Ising Hamiltonian, was defined as
\begin{equation}
h_{jk} = w_{h} h_{0} \left( \frac{D_{jk}}{D_\mathrm{ref}} \right)^{a_{jk}} \mathrm{cos} \left( \theta_{jk} \right)
\label{eq:37},
\end{equation}
where $h_0$ denotes the reference field-strength constant that determines the intrinsic energy scale of the local external-field term in the Ising Hamiltonian. The weight factor $w_h$  controls the relative contribution of the external-field term in the simulation. With this definition, cellular elements with higher local doses and greater alignment with the reference phase $\theta=0$ are subjected to a stronger external field, which preferentially drives their spins toward the reference $+1$ state.

\begin{figure}[b]
\includegraphics[width=0.95\columnwidth]{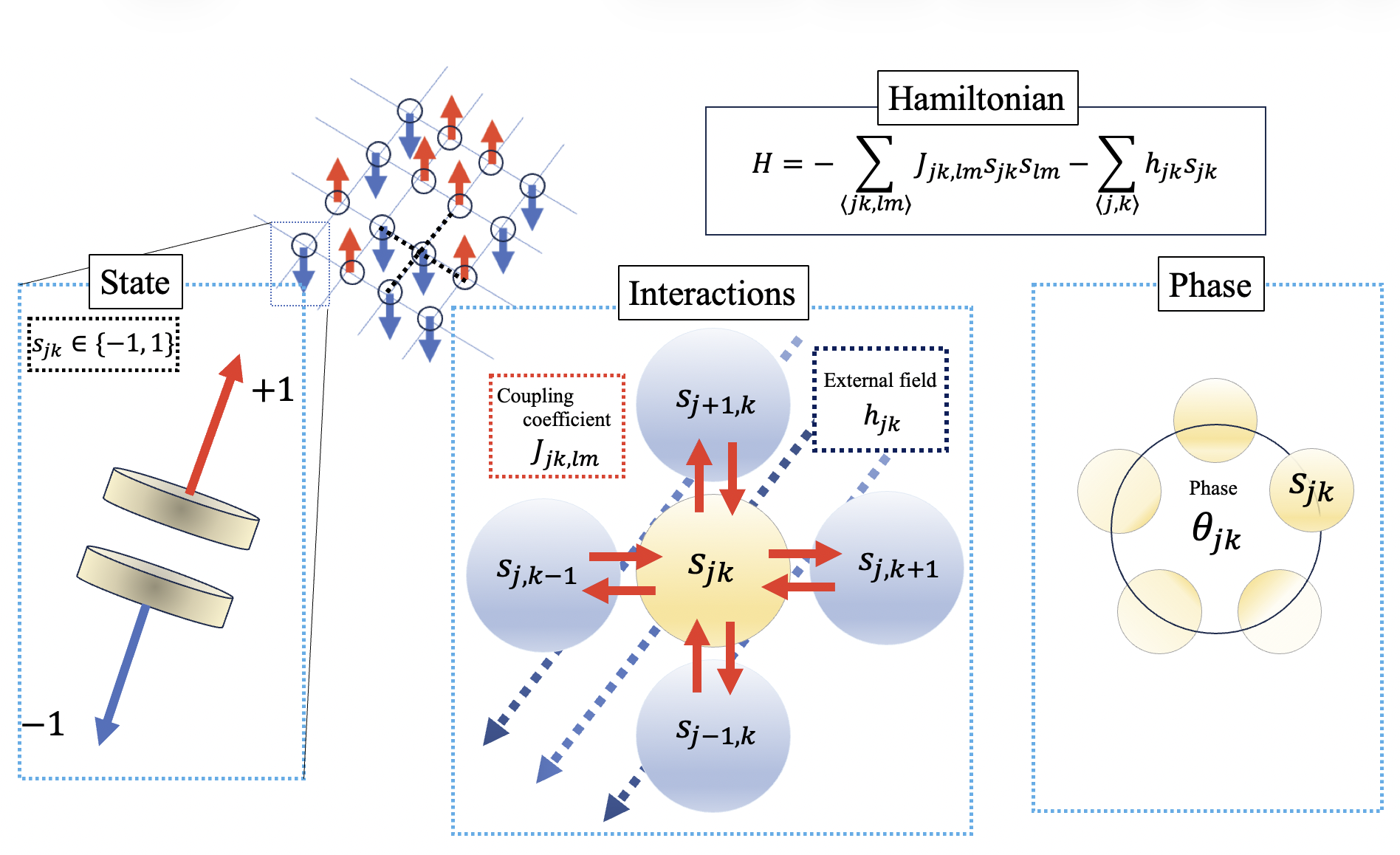}

\caption{\label{fig:epsart} 
Ising-model formulation in this study. Each cellular dose element is represented by an Ising spin variable $s_{j,k}\in\{-1,+1\}$, which encodes the local phase state or interaction polarity. The Hamiltonian consists of an interaction term between neighboring elements, governed by the coupling coefficient $J_{jk,lm}$, and a local dose-dependent external-field term $h_{jk}$. The phase parameter $\theta_{jk}$ modulates the interaction polarity among dose components, allowing phase-dependent collective effects to be incorporated into the QgEUD simulation.}

\end{figure}

The equilibrium spin configuration was estimated using the Metropolis algorithm. At each Monte Carlo step, a candidate cell $(j,k)$ was selected uniformly at random from all cells, and a spin flip $s_{j,k} \rightarrow -s_{j,k}$ was proposed. The energy change associated with this proposed flip was calculated from the corresponding change in the Hamiltonian ${\Delta H}$ as:
\begin{equation}
\Delta H =  H \left( s_{j,k} \rightarrow -s_{j,k} \right) -  H \left( s_{j,k} \right)
\label{eq:38}.
\end{equation}
The proposed flip was then accepted according to the Metropolis criterion,
\begin{equation}
\mathrm{min} \left( 1, \mathrm{exp} \left(-\frac{\Delta H\left( s_{jk} \right)}{T} \right) \right)
\label{eq:39},
\end{equation}
where $T$ denotes the simulation temperature. Thus, energy-decreasing updates $(\Delta H \leq 0)$ were always accepted, whereas energy-increasing updates $(\Delta H \geq 0)$ were accepted with probability $\mathrm{exp} \left( -\Delta H \left( s_{j,k} \right) /T \right)$. In this standard Metropolis formulation, the candidate cell is selected uniformly at random. Repeated application of this update rule generates a Markov chain whose stationary distribution is the Boltzmann equilibrium distribution, The acceptance ratio during the Metropolis update process was also monitored as an important indicator of simulation stability. In addition, the magnetization, defined as the global spin state of the system, was recorded at each step.

\subsection{\label{sec:level2} Calculation conditions}
In this study, an intensity-modulated radiotherapy (IMRT) dose distribution with a grid size of 2 mm was generated for a virtual phantom using the Eclipse treatment planning system (Varian Medical Systems, Palo Alto, CA, USA). Figure 3a shows the three-dimensional arrangement of the target and organs at risk in the virtual phantom. The clinical target volume (CTV) and organs at risk (OARs), including the bladder, rectum, and bilateral femurs, were delineated as regions of interest.
Figure 3b illustrates the computational workflow for QgEUD calculation. The region-of-interest (ROI) map was first converted into an initial phase map and an $a$-value map. The predefined organ-at-risk regions, collectively denoted as OAR1, were assigned $a=5$ or $a=8$ to represent serial-organ characteristics. The remaining internal body region, excluding these structures and denoted as OAR2, was assigned $a=3$ or $a=5$ to represent parallel-organ characteristics. The clinical target volume was assigned $a=-10$ or $a=-15$, reflecting the importance of minimum dose for tumor control. Random phase perturbations were then assigned to the OAR regions relative to the CTV phase. Subsequently, an Ising simulation was performed to estimate the equilibrium state arising from the local QgEUD affected by interactions and phase-dependent effects. Each simulation was annealed for 300 Metropolis Monte Carlo steps, and the resulting equilibrium spin configurations were averaged over $N$-independent simulations to obtain the expectation-value map. Finally, the expectation-value map was combined with the physical dose distribution using gEUD-based weighting to generate the final QgEUD-based biological-effect map. 
These biological effects were further evaluated using the Kähler geometric response, $\mathfrak{H}_K$. In this study, the QgEUD was evaluated on a cell-by-cell basis over the two-dimensional dose distribution. Accordingly, the gEUD kernel was discretized using the local dose $D_{jk}$ or the phase-affected dose $K_{jk}$ together with the corresponding $a$-value $a_{jk}$. From Eq.(12), the local geometric response is given by

\begin{equation}
\mathfrak{H}_{K_{jk}} =   \frac{1}{a_{jk}^2} \left| \frac{ \mathrm{log} K_{jk}}{a_{jk}} - \mathrm{log}D_{jk} \right|
\label{eq:40}.
\end{equation}
To evaluate the influence of the simulation parameters, the conditions of the Metropolis-based Ising model were systematically investigated. Unless otherwise specified, the weighting factors were fixed at $w_J=w_h=0.50$, and the reference coupling strength and external-field strength were set to $J_0=h_0=0.5$. The baseline simulation was performed using $T=0.8$, $(a_{\mathrm{CTV}}, a_{\mathrm{OAR}},a_{\mathrm{RVR}})=(-10.0, 5.0, 3.0)$, $N=250$, $J_0=0.5$, and $h_0=0.5$.

\begin{figure}[b]
\includegraphics[width=0.95\columnwidth]{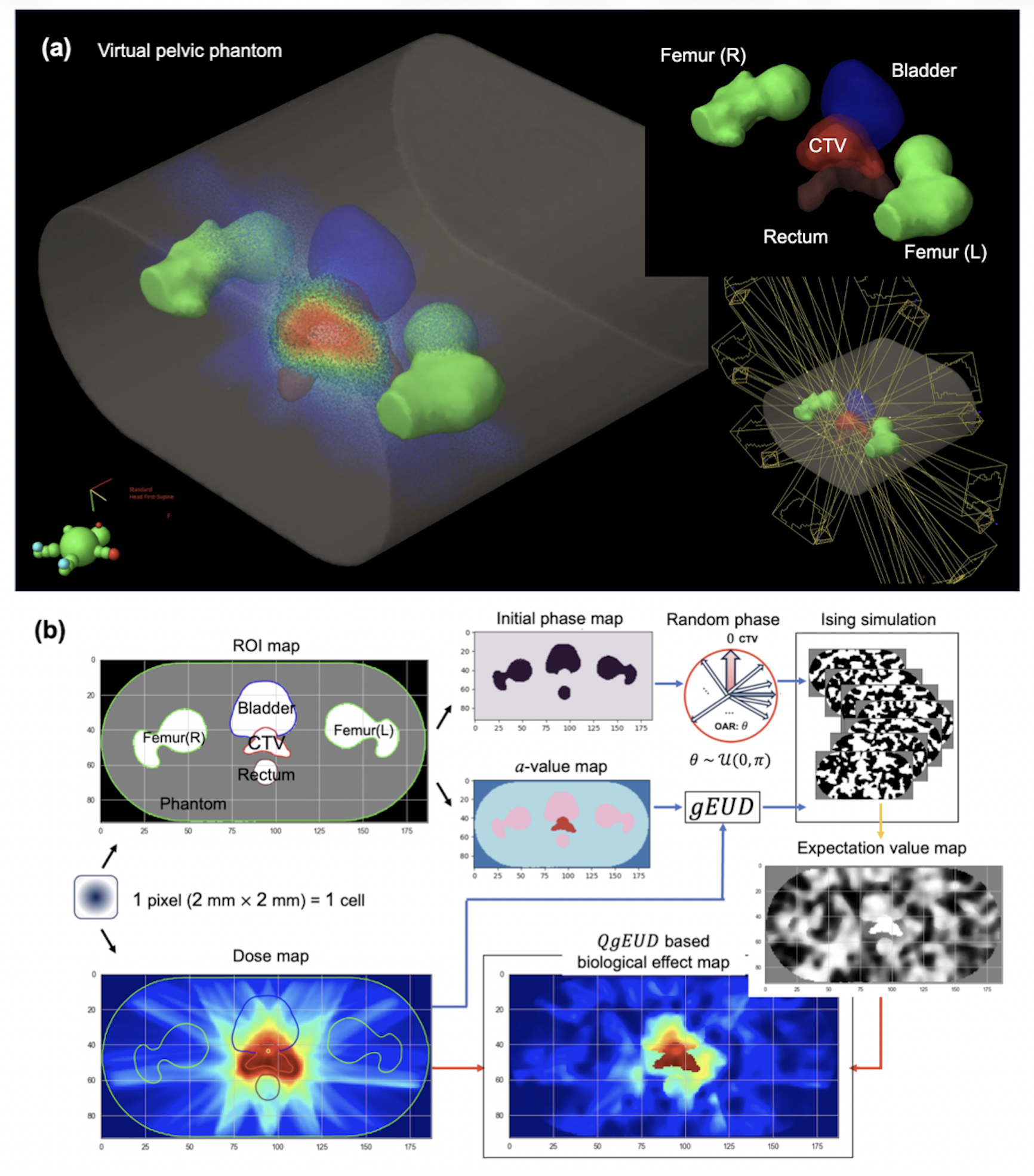}

\caption{\label{fig:epsart}
QgEUD calculation in a virtual phantom using Ising-model-based phase interaction analysis. (a) Spatial arrangement of the CTV and organs at risk in the virtual phantom. The bladder, rectum, and bilateral femurs were contoured, and the dose distribution was generated using intensity-modulated radiotherapy (IMRT), showing the lower-right inset the beam arrangement and spatial relationship between the dose distribution and organ structures. (b) Computational workflow for deriving the QgEUD map. The region-of-interest (ROI) and dose maps were first converted into an initial phase map and an $a$-value map. Random phase values, $\theta \sim U(0,\pi)$, were then assigned to non-target regions relative to the clinical target volume (CTV) phase. Ising simulation was subsequently performed to generate a series of spin configurations. The equilibrium expectation-value map was estimated by averaging the resulting spin distributions, thereby representing the collective effects of local interactions and phase coherence. Finally, the expectation-value map was integrated with the gEUD-based dose weighting to produce the QgEUD-based biological effect map.}
\end{figure}

\section{\label{sec:level1} Results \protect}
\subsection{\label{sec:level2} QgEUD-based Ising simulation}
The physical dose distribution generated by intensity-modulated radiotherapy is shown in Fig. 4a. The high-dose region was concentrated within the clinical target volume, whereas the surrounding organs at risk exhibited substantially lower dose levels. Using the assigned phase map and $a$-value map, the Metropolis-based Ising simulation produced the equilibrium expectation-value map shown in Fig. 4b. The expectation-value map exhibited spatially coherent response patterns reflecting the combined effects of local interactions and phase-dependent modulation, while preserving the overall structure of the original dose distribution.

The equilibrium expectation-value map was subsequently combined with the physical dose distribution using gEUD-based weighting to generate the QgEUD-based biological-effect map (Fig. 4c). Compared with the physical dose distribution, the resulting map emphasized regions in which both dose magnitude and phase-dependent interactions contributed to the biological response, while suppressing regions with relatively weak collective interactions.

Finally, the local Kähler geometric response was evaluated from the phase-affected QgEUD kernel on a cell-by-cell basis, yielding the map shown in Fig. 4d. Regions exhibiting strong dose-dependent biological effects generally corresponded to low geometric response, whereas regions with high geometric response were more strongly influenced by variations in the $a$-value, phase modulation, and intercellular interactions, indicating spatial heterogeneity in the geometric sensitivity of the QgEUD surface. The regions with elevated Kähler metric values were mainly distributed around the CTV-expanded area and along the surrounding dose-gradient region. This finding suggests that the local QgEUD response is amplified where the physical dose distribution, tissue-dependent $a$-value assignment, and phase-dependent interaction terms jointly vary. In such regions, small perturbations in dose weighting or phase modulation can induce relatively large changes in the QgEUD potential, suggesting an increased likelihood of radiobiological responses beyond those predicted from the physical dose distribution alone.

In the Ising simulation, the acceptance ratio decreased rapidly from approximately 0.70 at the beginning of the simulation and gradually stabilized at approximately 0.11–0.12 after 300 Metropolis Monte Carlo steps. Across 250 independent simulations, the late-stage acceptance ratio showed only small inter-run variability, suggesting that the spin configurations had reached a low-temperature stabilized regime without complete freezing (Fig. 4e). In parallel, the ensemble-averaged magnetization gradually increased from nearly zero to approximately 0.04 and reached a stable plateau in the late stage (Fig. 4f). The magnetization did not approach $\pm 1$, indicating that the spin field was not globally saturated. These results suggest that the Metropolis simulation reached a stabilized low-temperature regime while preserving local spin heterogeneity.

\begin{figure}[b]
\includegraphics[width=0.95\columnwidth]{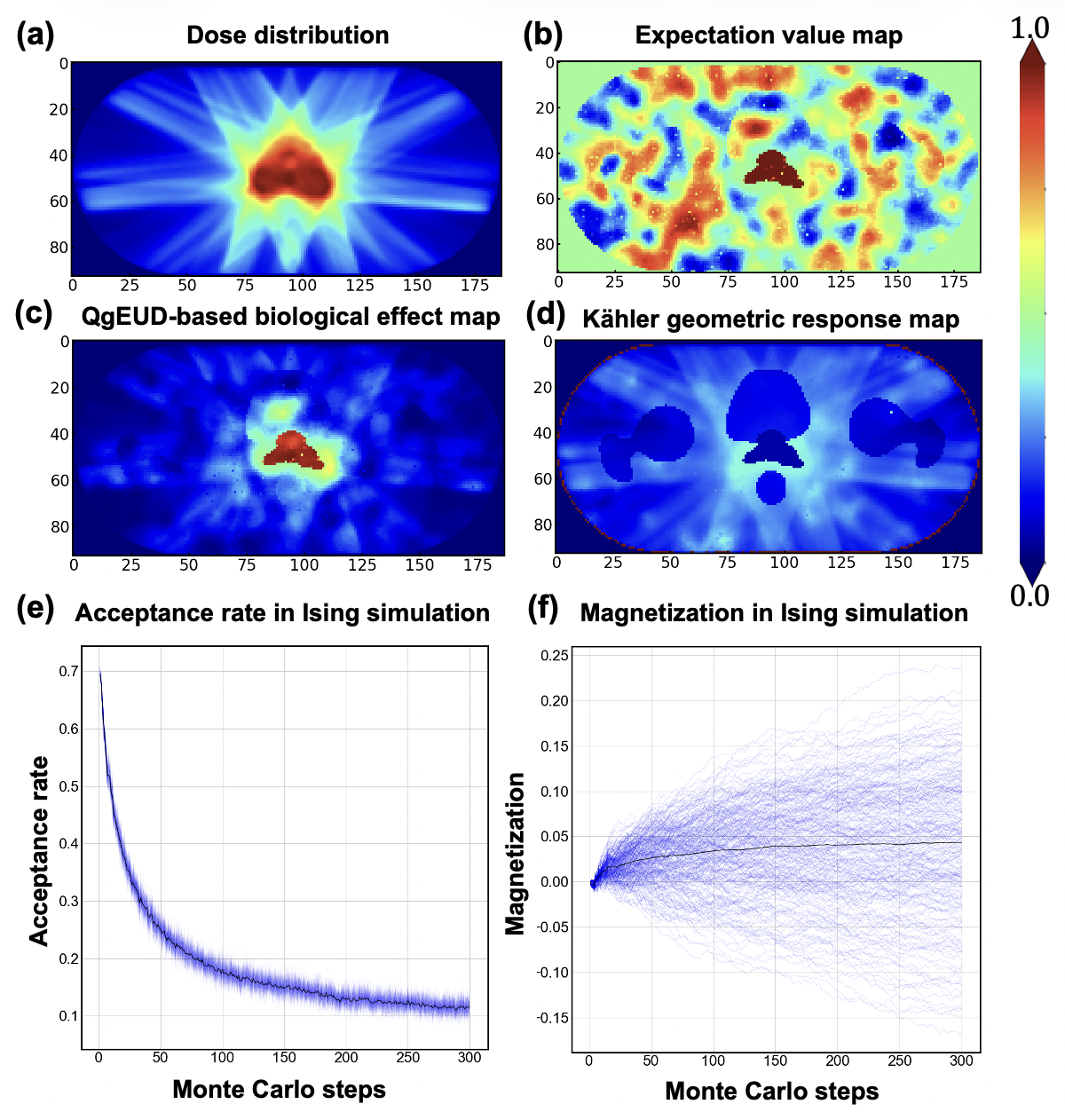}

\caption{\label{fig:epsart}
QgEUD-based biological-effect evaluation using the Metropolis-based Ising simulation. (a) Physical dose distribution generated for the virtual phantom. (b) Equilibrium expectation-value map obtained by averaging the spin configurations over 250 independent Metropolis-based Ising simulations. (c) QgEUD-based biological-effect map generated by combining the expectation-value map with the physical dose distribution using gEUD-based weighting. (d) Kähler geometric response map calculated from the local phase-affected QgEUD kernel, representing the local geometric response to perturbations in the $a$- and phase directions. (e) Evolution of the acceptance ratio during the Metropolis simulation, demonstrating convergence toward a stable equilibrium state. (f) Magnetization as a function of the Monte Carlo steps. (e)–(f) Blue curves represent individual simulation runs, and the black curve denotes the ensemble average, indicating stabilization of the global spin state during the annealing process.}
\end{figure}

\subsection{\label{sec:level2} Parameter sensitivity analysis of the metropolis-based Ising simulation}
To evaluate the influence of each parameter on the equilibrium spin configuration, one parameter was varied while the remaining parameters were fixed at their baseline values as shown in Fig. 5. The investigated parameter sets were as follows: 

\begin{equation*}
	\begin{split}
  		\bm{a}
		&= 
		\begin{pmatrix} a_{\mathrm{CTV}} \\ a_{\mathrm{OAR}} \\ a_{\mathrm{RVR}} \end{pmatrix}
  		= 
  		\begin{pmatrix} \{-15, -10 \} \\ \{ 5.0, 8.0 \} \\ \{3.0, 5.0\} \end{pmatrix},
	\end{split}
\end{equation*}

\begin{equation*}
	\begin{split}
  		T  &= \{ 0.4, 0.6, 0.8, 1.0, 1.2 \}, \\
  		N  &= \{ 50, 100, 250, 500, 1000 \}, \\
  		J_0 &= \{ 0.3, 0.4, 0.5, 0.6, 0.7 \}, \\
  		h_0 &= \{ 0.3, 0.4, 0.5, 0.6, 0.7 \}.
	\end{split}
\end{equation*}

Increasing the number of independent simulations from $N=50$ to $N=250$ progressively stabilized the expectation-value map, whereas further increases to $N=500$ and $1000$ produced only minor visual changes. The simulation temperature had the largest influence on the spatial distribution. At low temperatures $T=0.4$ and $0.6$, high-response clusters were strongly preserved, whereas higher temperatures $T=1.0$ and $1.2$ reduced the contrast and weakened local structures. The reference temperature $T=0.8$ preserved the global dose pattern while maintaining distinct local features. Variation of the gEUD parameter set produced relatively moderate changes compared with those of $T$ and $J_0$. Among the investigated parameter sets, $\bm{a}=(-10,5,3)$ preserved the overall spatial dose-response pattern without excessive emphasis on either low- or high-dose regions.

The reference coupling strength $J_0$ markedly affected spatial coherence. A low coupling strength $J_0=0.3$ resulted in fragmented response patterns, whereas higher values $J_0=0.6$ and $0.7$ increased the connectivity and aggregation of high-response regions. The reference value $J_0=0.5$ maintained coherent local structures without excessive clustering. In contrast, variation of the reference external-field strength $h_0$ produced comparatively small changes in the global response pattern. Increasing $h_0$ slightly enhanced the contrast of high-response regions while preserving the overall spatial distribution. Based on these observations, the parameter set highlighted in Fig. 5, $N=250$, $T=0.8$, $\bm{a}=(-10,5,3)$, $J_0=0.5$, and $h_0=0.5$, was adopted as the baseline condition for subsequent QgEUD calculations because it preserved the global dose structure while maintaining locally modulated response patterns.

\begin{figure}[b]
\includegraphics[width=0.95\columnwidth]{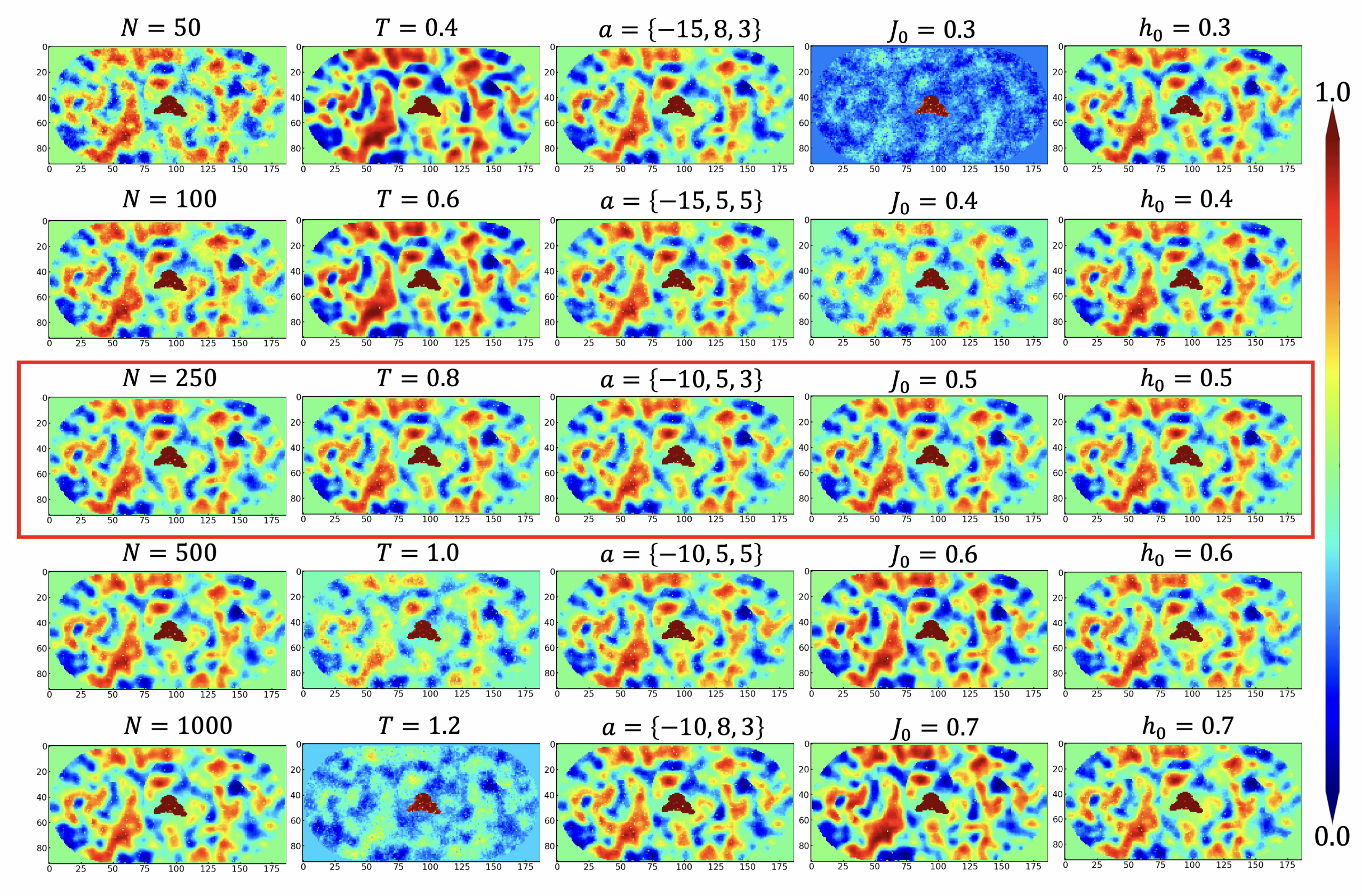}

\caption{\label{fig:epsart}
Parameter sensitivity analysis of the Metropolis-based Ising simulation. The effects of the number of independent simulations $(N)$, simulation temperature $(T)$, gEUD factor $(a)$, reference coupling strength $(J_0)$, and reference external-field strength $(h_0)$ on the equilibrium expectation-value map normalized from $0$ to $1$ are shown. Increasing $N$ improved convergence of the spin configuration, whereas $T$ controlled the balance between stochastic fluctuations and spatial ordering. The parameter $a$ determined the relative sensitivity to low-, intermediate-, and high-dose regions, while $J_0$ and $h_0$ governed the strengths of intercellular interactions and local dose-dependent bias, respectively. The red rectangle indicates the baseline parameter set used throughout this study $\left( N=250, T=0.8, \bm{a}=(-10,5,3), J_0=0.5, h_0=0.5 \right)$. Under these conditions, the global structure of the physical dose distribution was preserved while local interaction-driven modulation was maintained without excessive smoothing or stochastic noise, and this parameter set was therefore adopted for subsequent QgEUD calculations.}
\end{figure}


\section{\label{sec:level1} Discussion \protect}
\subsection{\label{sec:level2} Conceptual significance of QgEUD}
This study proposed the quantum generalized equivalent uniform dose (QgEUD) as a phase-extended formulation of conventional gEUD and demonstrated its feasibility using a Metropolis-based Ising simulation. Unlike conventional gEUD, which summarizes heterogeneous dose distributions solely according to dose magnitude, QgEUD incorporates an additional phase dimension to represent interaction-dependent biological responses. Consequently, dose-derived amplitude and phase-dependent modulation are treated within a unified mathematical framework.

This extension provides three conceptual advantages. First, QgEUD enables biological effects to be evaluated not only by dose magnitude but also by relative phase relationships among cellular dose components, allowing enhancement, suppression, and interference-like responses to be represented. Second, the QgEUD manifold provides a geometric interpretation of dose-response sensitivity. The conventional gEUD curve is recovered on the real section $\theta=0$, whereas phase modulation extends the curve into a two-dimensional manifold, allowing local sensitivity to be evaluated in both the dose-weighting and phase directions. Third, the formulation is naturally compatible with Ising and QUBO representations, enabling interaction-aware dose evaluation within a quantum-inspired computational framework.

\subsection{\label{sec:level2} Interpretation of the QgEUD response maps}
The proposed framework preserved the global structure of the physical dose distribution while introducing local modulation through the equilibrium expectation-value obtained from the Ising simulation. Rather than replacing the physical dose distribution, the QgEUD response map can therefore be interpreted as a phase- and interaction-weighted modulation of conventional dose evaluation.
The expectation-value map represents the equilibrium tendency of neighboring cellular dose elements under the combined influence of dose similarity, phase alignment, and local external fields (Fig. 4b). When combined with the physical dose distribution through gEUD weighting, it yields a biologically interpretable response map that bridges voxel-wise physical dose evaluation and population-level collective response.

The resulting QgEUD map preserved the overall dosimetric pattern while enhancing the biological response around the CTV (Fig. 4c). In contrast, the Kähler geometric response did not simply reproduce the physical dose distribution but instead highlighted transition regions surrounding the target and high-dose-gradient areas (Fig. 4d). This finding indicates that the Kähler metric captures phase- and interaction-dependent sensitivity beyond dose magnitude alone, thereby identifying regions where biological responses may deviate from those predicted by conventional dose-based evaluation. Consequently, the Kähler geometric response provides complementary information to QgEUD by localizing regions in which collective cellular interactions and phase modulation are expected to play a dominant role, potentially offering additional guidance for adaptive radiotherapy and biologically informed treatment optimization.

\subsection{\label{sec:level2} Validation of the proposed simulation}
The convergence behavior of the Metropolis simulation supports the internal consistency of the proposed framework. The acceptance ratio stabilized at approximately 0.1–0.2 without collapsing to zero (Fig. 4e), indicating that the spin system approached a stable equilibrium while maintaining sufficient configurational exploration \cite{refs21}. Similarly, the ensemble-averaged magnetization converged to a small positive value rather than approaching complete spin saturation (Fig. 4f), indicating preservation of local heterogeneity.

Parameter sensitivity analysis further demonstrated that the adopted baseline condition provided a balanced compromise between convergence, spatial coherence, and preservation of the original dose structure (Fig. 5). The simulation temperature and coupling strength had the greatest influence on local response patterns, whereas increasing the number of independent simulations improved the stability of the expectation-value map with only marginal changes beyond N=250. Together, these observations support the robustness of the selected simulation parameters.

\subsection{\label{sec:level2} Relationship to existing radiobiological models}
QgEUD extends existing dose-response models rather than replacing them. Conventional DVH metrics, gEUD, TCP, and NTCP primarily describe biological response as a function of dose magnitude and tissue-specific volume effects. By introducing phase as an additional latent variable, QgEUD provides a mathematical representation of collective interactions that are not explicitly considered in conventional approaches.

Similarly, whereas many radiobiological models explicitly model oxygenation, DNA repair, or cell-cycle kinetics, QgEUD represents these coupled effects abstractly through phase modulation rather than mechanistic biological parameters. From the perspective of quantum information science, the present framework should be regarded as quantum-inspired. Although mathematically compatible with Ising and QUBO formulations for quantum annealing and QAOA, all simulations in this study were performed using a classical Metropolis algorithm. Nevertheless, the expectation-value formulation of QgEUD suggests a possible route toward future quantum implementations, in which techniques such as the Hadamard test could be employed to quantify phase-dependent radiobiological effects directly from quantum states \cite{refs22}. This perspective extends QgEUD beyond a classical simulation framework toward a quantum-measurable representation of biological response.

\subsection{\label{sec:level2} Limitations and future perspectives}
Several limitations should be acknowledged. First, the present study was performed using a two-dimensional virtual phantom and coarse-grained cellular elements rather than true biological cells. Extension to three-dimensional patient dose distributions and multiscale biological modeling will therefore be necessary.

Second, the phase parameter was introduced as a random perturbation relative to the CTV phase. Future work should relate phase to measurable biological quantities, such as oxygenation, linear energy transfer (LET) distribution, DNA damage response, or intercellular signaling.

Third, the parameters used in the present study, including the $a$-values, coupling strength, external-field strength, and simulation temperature, were selected empirically. Their biological interpretation should ultimately be established through calibration against experimental or clinical outcome data.

Finally, although the proposed framework is compatible with quantum computing, practical implementation on current quantum hardware remains challenging because direct pixel-wise encoding of clinical dose distributions would require an impractically large number of qubits. Future work should therefore investigate compressed encodings, ROI-level representations, hybrid classical–quantum algorithms, and quantum annealing strategies.

Overall, the present study demonstrates that QgEUD extends conventional gEUD by incorporating phase-dependent modulation into dose aggregation while preserving the underlying physical dose distribution. The proposed framework provides biologically interpretable response maps, Kähler geometric sensitivity, and a computational structure suitable for future quantum-inspired optimization. Although further biological validation and quantum implementation remain necessary, QgEUD offers a promising theoretical framework for integrating heterogeneous dose distributions, collective cellular interactions, and phase-dependent radiobiological effects within a unified dose-response model.

\section*{Ethical approval}
The study protocol was reviewed and approved by the Institutional Review Board and Independent Ethics Committee of Kansai Medical University (approval no. 2022065). Informed consent was obtained through an opt-out procedure, and individuals who declined participation were excluded from the study. All procedures were performed in accordance with the relevant guidelines and regulations, the principles of the Declaration of Helsinki, and applicable local statutory requirements.

\section*{No conflict of interest}
The author declares that there are no conflicts of interest to disclose.

\section*{Data availability statement}
Derived data and calculation code supporting the findings of this study are available from the corresponding author Y. A. on request.

\section*{Acknowledgments}
This work partly supported by [grant numbers 18K15650, 25K19118] from the Japan Society for the Promotion of Science.

\clearpage
\section*{}
\typeout{}
\bibliographystyle{unsrt}
\bibliography{MS_QgEUD_Anetai_arXiv_bib}

\end{document}